\documentstyle[psfig]{caosp}

\begin{document}
\pubyear{1998}
\volume{27}
\firstpage{272}
\htitle{Theoretical aspects of roAp stars}
\hauthor{M. S. Cunha}
\title{Theoretical aspects of roAp stars}
\author{M. S. Cunha}
\institute{Institute of Astronomy, Madingley Road, Cambridge CB3 0HA, U.K.}
%\date{}
\maketitle

\vspace{-0.2cm}
\section{Introduction}
\vspace{-0.2cm}

Since rapidly
oscillating Ap stars (roAp) were first discovered (Kurtz 1982), the number  
observed has increased considerably, making today a total of 28. This discovery
of new roAp
stars, together with better observations of the ones already known, have
brought
to light many interesting questions, revealing, at the same time, the need for
further theoretical studies on the subject.

Among the many observational facts that need to be understood (for a review on
the observational facts of roAp stars see Kurtz, 1990, 1995) are the high
frequencies of the modes observed, which can be higher than the
theoretical critical cutoff frequency for acoustic modes in these stars, their
apparent alignment with the magnetic field, and the fact that some modes
cannot be described by one single spherical harmonic.

In this paper the theoretical work on the roAp stars will be reviewed, and
the implications of this work on the questions mentioned above will be
inspected. In section 2 the different mechanisms
proposed to excite pulsations in these stars will be described, and related
to the high frequencies of the modes, and their alignment with the magnetic
field. In section 3 the methods commonly used to infer
information about these stars, from the observation of their oscillations,
will be reviewed,
and the problems associated with these methods, in particular
when the magnetic field is taken into account, will be discussed.
\vspace{-0.1cm}
\section{Excitation mechanisms}
\vspace{-0.2cm}

\subsection{$\kappa$-mechanism}

In the HR diagram the roAp stars are located right in the instability
strip for classical pulsators, like Cepheids, RR Lyrae and $\delta$-Scuti,
and, since
the pulsations in the latter are known to be driven by the $\kappa$-mechanism,
it was soon proposed that this should be the mechanism exciting the
oscillations in
the roAp stars. There is, however, one important difference between the
oscillations
in the classical pulsators and those in the roAp stars, which is their
periods.
The $\delta$-Scuti, for instance, which, among the classical pulsators,
are the closest in luminosity to the roAp stars, have periods between
0.02 and 0.25 days, in contrast with the typical periods of the roAp stars, 
which range between 5.6 and 15.0 minutes.
This difference in the periods of oscillations has severe consequences for the
excitation process proposed here, since the efficacy of the
$\kappa$-mechanism depends largely
on the periods of the modes.

The $\kappa$-mechanism can drive pulsations in the
regions of the star where the opacity increases during contraction of the
material.
That happens particularly in regions in which the gas is partially ionized.
Associated with each ionization region is a different thermal relaxation time
scale, and only the modes
with periods similar to one of these time scales can be excited by
the corresponding layer.
Consequently, if it is indeed the $\kappa$-mechanism which is
responsible for
exciting the pulsations observed in the roAp stars, then the region  in the
star where these modes are
excited must be different from that where the excitation of the modes
observed in $\delta$-Scuti stars
takes place, as the frequencies of the modes in these stars are so different.
In the case of $\delta$-Scutis, the modes are known to be excited
in the
region of second ionization of helium. In roAp stars, however, the period
of the oscillations is of the order of the thermal relaxation time only in
the region of ionization
of hydrogen. It is therefore expected to be this layer where most of
the driving takes place.

But, for the oscillations to be excited by the $\kappa$-mechanism, there is
still another
condition that must be fulfilled, namely that the energy gained by the
oscillation in the regions where
the $\kappa$-mechanism is exciting it overcomes all the energy losses that
take place
throughout the star.
Calculations (Dziembowski \& Goode, 1996) indicate that in Ap-star models
with chemically homogeneous
envelopes only low overtones are excited by the $\kappa$-mechanism.
 So, in fact, two questions must be posed at this point:
first, why do roAp stars pulsate in high frequencies and, second, why do
they not pulsate at low
frequencies, like the classical pulsators?
The main idea proposed to answer both questions goes as follows
(Dolez \& Gough 1982,
Dziembowski \& Goode 1996):  Imagine that, in these stars, helium settles
by gravity, while hydrogen
rises to take its place. If this were to be the case, there would be a
depletion of helium in the region of the second ionization
of helium, justifying the absence of low overtone pulsations, while an
excess of hydrogen would be accumulated
in the region of partial ionization of hydrogen, resulting in a greater
efficacy of the
$\kappa$-mechanism in this layer and, possibly, exciting the high
overtones. If, together with this,
the settling of helium were to take place preferentially around the magnetic
poles, rather than near the
equator, then the driving of pulsations would take place around the poles,
and the modes would be
excited preferentially in alignment with the magnetic field, as it is
observed. But, if this
is really the clue for the excitation of pulsations in roAp stars, then we
are left with two new questions,
namely of whether this settling  of helium should take place, and why
should it be more efficient at the
poles?

If the shallow convective layer present in these stars is sufficient to
prevent the settling of helium,
then one possible answer would be that the magnetic field could suppress
convection around the magnetic
poles, allowing for helium to settle in this region, while in the rest
of the star the mixing
of helium would take place normally. There is, however, some discussion of
whether the magnetic field would
suppress convection only at the poles, and, maybe even more important, of
whether the convective layer
in these stars can efficiently  mix helium at all. In any case, since
the distribution of the chemical
composition in these stars is clearly linked with the magnetic field, as is
seen from the formation of
spots around the magnetic poles, it seems plausible that a preferential
settling of the helium, linked with the magnetic field, could
take place. Following these ideas,
Dolez \& Gough (1982) carried out calculations for the growth rates of high
overtones in a stellar
model where, at the magnetic poles, the convection was made less efficient
than in the usual models and
all the helium above the convective layer was replaced by hydrogen.
Unfortunately,
no excitation of high overtones was found. It is well known,
however, that the opacity tables have changed since 1982, and this might
change the results.
But that work is still to be done.

As said before, one fundamental requirement in order to excite oscillations
is that the energy
gained by the mode, in the regions where it is being excited, overcomes the
energy lost due to all
dissipation processes.
One of such processes, through which energy is lost, takes place when the
modes propagate all
the way to the outermost layers of the star, without ever being reflected.
Only modes that are
reflected by the atmosphere, and, therefore, trapped between two turning
points, can accumulate enough
energy and grow to observable amplitudes.
Since roAp stars pulsate in frequencies above the theoretical critical
cutoff frequency for acoustic
modes in these stars, the question of how is it possible for these modes to
be trapped, and, therefore,
observed, has soon been posed. Our poor knowledge of the atmosphere of
these stars, and in particular
of the gradient of temperature, might be partially responsible for the apparent
incompatibility between the critical cutoff frequency and the frequency of
the modes observed. There is,
however, something that cannot be forgotten when posing this problem, and
that is the magnetic
field, which has both direct and indirect effects on the oscillations
(Shibahashi \& Saio 1985).
The direct effect of the magnetic field will be briefly discussed.
It is well known that the surface magnetic field in these stars is
large and that
in the outer layers the pulsations are largely affected by this field.
In particular, the modes
will be magnetoacoustic in nature, and, therefore, the ability of the
atmosphere to reflect them
should no longer be associated with the value of the critical cutoff
frequency for acoustic modes.
This fact is well illustrated in the simple case of a plane parallel,
isothermal atmosphere, with
constant gravity and with an horizontal magnetic field (Stark \& Musielak
1993). The critical cutoff
frequency changes in this model, from its usual value of
$\omega_c=\frac{c}{2H}$, when there is no
magnetic field, to
$\omega_{\rm ac}=\left[\omega_{\rm c}^2+\frac{3}{4}\frac{V_{\rm A}^2}
{\left({c}^2+{V_{\rm A}}^2\right)
}
\left(\frac{V_{\rm A}}{2H}\right)^2\right]^{1/2}$, when the magnetic field is
turned on, where
$c$ is the sound speed, $V_{\rm A}$ is the Alfv{\'e}n speed and $H$ is the
density scale height.
In this particular case, the presence of the magnetic field has very strong
implications concerning
the reflection of the modes, since $V_{\rm A}$ tends to infinity as the density
decreases towards zero, and,
consequently, the critical cutoff frequency tends itself to infinity,
guaranteeing the reflection
of modes of any frequency. In the roAp stars the magnetic field is thought
to be approximately
dipolar, and the study of reflection of the magnetoacoustic waves,
therefore, grows in complexity.
In any case, it is clear that such study should be carried out before
assuming that the frequency
of the modes observed in these stars represents a problem.

\subsection{Magnetic overstability}

Soon after the discovery of the roAp stars, an alternative mechanism was
proposed to drive pulsations
in these stars (Shibahashi 1983, Cox 1984): the magnetic overstability. The
main idea behind this
mechanism is that,
if we impose a magnetic field on a superadiabatic layer, the motion, which
would normally be convective,
can become oscillatory. This happens because the putative convective
motion distorts the magnetic field
lines, which, in turn, react back through the form of a magnetic force,
which opposes the original motion.
The ability of the magnetic force to reverse the direction of the motion
depends on the strength of
the magnetic force, when compared with the buoyancy force. Only if
the magnetic force is greater than the buoyancy force
does the oscillatory motion  replace the convective motion. If, together
with satisfying this condition, there is a mechanism to make this
perturbation grow with time,
then modes will be generated that can potentially be observed. These will
be transverse modes,
usually called magneto-gravity modes.
To understand how the magneto-gravity modes can grow with time, the
nonadiabatic processes that
take place during this motion must be taken into account. Throughout the
motion there
is heat exchanged between the eddy and the surroundings, which is
associated with the thermal diffusivity
of the fluid; but there is also a magnetic diffusivity, related with the
slippage of the
magnetic field lines through the fluid. If the magnetic diffusivity is
smaller than the thermal diffusivity,
then the buoyancy force decreases with time faster than the magnetic force,
and, consequently, the
total restoring force acting on the fluid increases with time. This
increase results in the
amplification of the mode, which can, in this way, grow to observable
amplitudes. Otherwise, the oscillations decay. Under the conditions
present in these stars, in the superadiabatic layer, the magnetic
diffusivity is much smaller
than the thermal diffusivity, and, therefore, the conditions needed to
drive the magneto-gravity
modes are well satisfied.

For reasonable values of density,
magnetic field and wavenumber, Shibahashi calculated a frequency of about
1.6 $\mu$Hz for the pulsations,
which is well within the observed values.
Moreover, the magnetic restoring force is proportional to ${\bf B\cdot k}$,
where ${\bf B}$ is the magnetic field and ${\bf k}$ is the wavenumber,
 and, therefore, assuming the magnetic
field to be dipolar, and having in mind that the observed modes are of low
degree, for which
the vertical wavenumber is much greater that the horizontal wavenumber, one
concludes that
the motion will be transformed into oscillatory motion preferentially
around the poles, since
it is at the poles that ${\bf B\cdot k}$ takes its maximum value. This, in 
turn,
results in preferential driving of those modes aligned with the magnetic 
field.

Although the magnetic overstability seems to explain the frequencies of the
modes, and their
alignment with the magnetic field, in a natural way, it should be noticed
that this theory is
in fact quite incomplete, in the sense that only the excitation layer was
treated during the
analysis. As said before, in order for the modes to be excited, the energy
gained through this
mechanism has to overcome all the energy losses throughout the star. This
balance of energy
has never been investigated for modes excited in this way, so it is not
possible to conclude,
from the analysis done, whether the oscillations in roAp stars can in fact be
excited through this mechanism.

\vspace{-0.2cm}
\section{Asteroseismology and the magnetic field}
\vspace{-0.2cm}

In a spherically symmetric star, the frequency of high order p-modes is
asymptotically given
by the well known expression (Tassoul 1980):
\begin{equation}
\nu_{n,l}=\nu_0\left(n+\frac{l}{2}+\epsilon\right)-\frac{\left[l\left(l+1\right)
+\epsilon\right]
A_0\nu_0^2}{\nu_{n,l}} +\epsilon,
\end{equation}
where $\nu_{n,l}$ is the frequency of a mode of order $n$ and degree $l$,
$\nu_0=\left(2\int_0^R{\frac{{\rm d}r}{c}}\right)^{-1}$ and $\epsilon$ and
$A_0$ are constants
that also depend on the stellar structure.

In practice, when the amplitude spectrum of a star is given, the
information is organized
in two parts: the large separations, given by ($\nu_{n,l}-\nu_{n-1,l}$), and
the small separations, usually defined as ($\nu_{n,l}-\nu_{n-1,l+2}$).
The large separation scales roughly as $\left(\frac{M}{R^3}\right)
^{\frac{1}{2}}$, and, therefore,
gives information about the mass-radius relation of the star, which, with
the appropriate stellar model,
can be used to calculate the stellar luminosity. The small separation, on
the other hand,
depends on the gradient of the sound speed and, hence, is sensitive to the
core of the star.

So, if the modes observed in roAp stars are in fact acoustic, and if their
degrees can positively
be identified, then the large and small separations can be determined, and,
in principle,
information about the star can be obtained. There are, however, two
conditions that I have mentioned,
which are not always fulfilled: first, the equilibrium state of these stars
cannot be spherically
symmetric, due to the presence of the magnetic field and of chemical
inhomogeneities, and,
second, the modes observed in these stars cannot always be identified,
and, some times, cannot
even be described by a single spherical harmonic. So, these are two
problems that must be kept
in mind when trying to do asteroseismology with the roAp stars, and they
will be given some
attention in the following paragraphs.

The problem of identification of the modes in roAp stars is probably best
exemplified by
the well-studied star HR1217 (Kurtz et al 1989). For many years the question
has been raised
of whether the modes observed in this star were all dipole modes with
consecutive orders,
or if, instead, they were alternating even and odd degree modes. Depending
on which of these
interpretations is right, the values obtained for the large separations are
very different,
and so are the radii and luminosities determined for the star. With
the knowledge of HIPPARCOS paralaxes this problem has been solved (Kurtz
1998), in favour
of the second solution. However, this particular example shows how
important is the
correct identification of the modes, in order to infer correct information
about the star.

%%%%%%%
\begin{figure}
\centerline{\psfig{figure=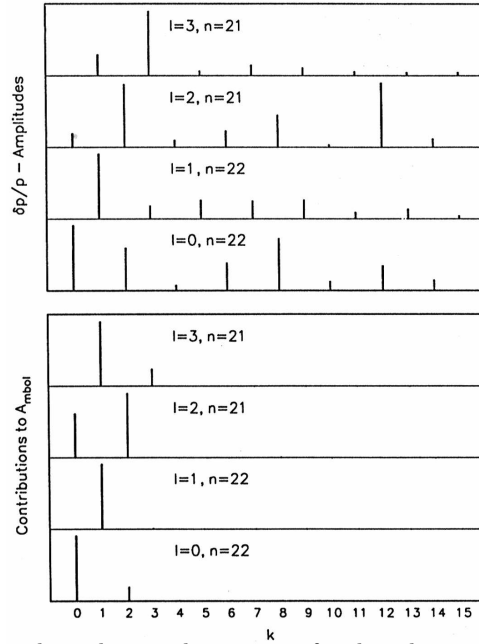,width=6.5cm}}
%\vspace{-0.3cm}
%\centerline{\psfig{file=pfig3.eps,width=8.5cm,height=8.5cm}}
\vspace{-0.2cm}
\caption{{\bf top:} Legendre polynomial expansion for the relative
pressure amplitude at the photosphere
for four modes of an Ap star model with {\it M}=2.0$M_{\odot}$,
{\it R}=2.138$R_{\odot}$, log $T_{\rm eff}$=3.9240,
log {\it L}/$L_{\odot}$=1.307 and $X_c$=0.38. {\bf bottom:} Contribution to 
bolometric amplitude variation calculated
under the assumption that the bolometric intensity follows the angular
dependence of the relative
pressure amplitude. The results are averaged over random orientations of
the axis of the magnetic
field.}
\vspace{-0.1cm}
\end{figure}
%%%%%%%

In roAp stars the modes are usually identified through the inspection of
the amplitude spectrum.
This inspection, however, does not take into account the fact that these
stars have strong
magnetic fields, which are likely to distort the eigenfunctions. In fact, the
magnetic field can distort
the eigenfunctions in such a way as to lead to the wrong identification of
the modes. This fact can
be seen in the work developed by Dziembwoski \& Goode (1996), where
they calculated the effect of a dipolar magnetic field on the
eigenfunctions. To do so, they
expanded the perturbed eigenfunctions as a linear sum of Legendre
polynomials of different degree.
The results obtained are shown in Fig. 1 (top) for the relative pressure
amplitude at the
photosphere, k being the degree of the Legendre polynomial in the
expansion, and l and n being,
respectively, the degree and order of the unperturbed mode. It is clear from
this plot that
modes that were originally characterized by a well determined degree l, are
distorted in the presence
of a magnetic field, since they show nonzero components of other degrees
in the expansion. In
other words, these modes will not be observed as single spherical
harmonics,  but as the sum of
several spherical harmonics of different degrees. Moreover, recalling that
what is observed
is, in fact, an average of the perturbations over the whole disk, and that,
in this way, the higher
degree components are averaged out, it is concluded that modes with higher
degrees might be wrongly
identified as having lower degree. This can be seen in Fig. 1 (bottom),
where the contributions to
the luminosity are given for the same model, after averaging over the disk.
From this plot it
is clear that, in general, when observed, the modes will be distorted from
single spherical harmonics.
In particular, the unperturbed l=3 mode presents, in the presence of the
magnetic field, and after
averaging over the disk, a dipole component that is stronger than the
original l=3 component, and might,
therefore, be misidentified as being a dipole mode.

The second problem that was mentioned relates with the nonspherical symmetry
of the equilibrium
 state, which is a consequence of the nonspherical symmetry of the
 magnetic field and chemical inhomogeneities present in these stars. These
are both surface effects:
 the chemical
inhomogeneities, because they are supposed to be present only at the
surface, and the magnetic field,
because it is only at the surface that the magnetic stress is comparable
with the pressure forces.
One question that is important to raise is how are the large and small
separations influenced
by these surface effects? This problem was investigated by several authors
(Balmforth et al. in preparation, Dziembowski \& Goode 1996,
 Cunha \& Gough 1998), and the conclusions were
that the small separations are largely affected by these surface effects,
and, consequently,
cannot be used to infer reliable information about the stars, while the
large separations
are influenced in a systematic way, but by an amount which is small when
compared with the large separations themselves.

There is still some hope, however, that in the future the small separations
might be used to infer information about these stars, if only modes with
nonzero azimuthal order, $m$,
can be observed. This happens because, if for each degree $l$ all the
$2l+1$ modes with different $m$
were observed, then, their arithmetic average would give the frequency of
the modes as if the
star were spherically symmetric (Gough 1993).

\vspace{-0.2cm}
\section{Discussion}
\vspace{-0.2cm}

As discussed in the first section of this paper, two main mechanisms have been
proposed to excite
oscillations in the roAp stars: the $\kappa$-mechanism and the magnetic
overstability.
These mechanisms are completely different in nature, and so are the
oscillations excited by
each of them. While the $\kappa$-mechanism drives acoustic modes, the
magnetic overstability
drives magneto-gravity modes, which are transverse.
It should be clear, however, that none of these mechanisms solves, at the
present time, the
problem of excitation of oscillations in roAp stars, since, for the
magnetic overstability there are no
calculations of the
growth rates, and, for the $\kappa$-mechanism, the growth rates obtained
were negative, meaning
no excitation. However, these last calculations should be repeated using
the most recent opacities.

Moreover, the recently confirmed success (Kurtz 1998) in the determination
of luminosities of roAp
stars, using the asymptotics for high-order acoustic modes, is a very
strong evidence against
the magnetic overstability, since, if the modes were excited through this
mechanism, they
would not be acoustic modes, and, therefore, the asymptotics would not apply.

Finally, in relation to the critical cutoff frequency, it was concluded
that the magnetic field cannot be separated from the problem of the reflection
 of the high frequency
modes by the atmosphere,
as these modes are in fact magnetoacoustic oscillations.

As for the asteroseismology in roAp stars, it seems clear that some care
must be taken. First,
 the identification of the modes might be a problem, since the magnetic
field tends to distort the
eigenfunctions. Secondly, the small and large separations are influenced by
non-spherically-symmetric
surface effects, like the magnetic field and chemical inhomogeneities. In
this case, the greatest
problem relates to the small separations, from which no reliable
information can be obtained.

%\vspace{-0.2cm}

\acknowledgements
MC acknowledges the support of JNICT (Portugal) through a
grant BD/5519/95, PRAXIS XXI programme.

\vspace{-0.2cm}

%{\Large \bf References}
%\vspace{0.2cm}

%{\small

\end{document}